\documentclass[amsmath,amssymb,prl,aps,twocolumn,mathbbm,superscriptaddress]{revtex4-1}
\usepackage{graphicx,dsfont}
\usepackage{epsfig}
\usepackage{physics}
\usepackage[normalem]{ulem}
\usepackage[dvipsnames]{xcolor}
\usepackage[colorlinks=true,citecolor=Cerulean,linkcolor=RubineRed,urlcolor=Cerulean]{hyperref}

\begin{document}
\title{Enhanced force sensitivity and entanglement in periodically driven optomechanics}
\author{F. Cosco, J. S. Pedernales, and M. B. Plenio}
\affiliation{Institut f\"ur Theoretische Physik und IQST, Albert-Einstein-Allee 11, Universit\"at Ulm, D-89081 Ulm, Germany}

\begin{abstract}

Squeezing is a resource that enables precision enhancements in quantum metrology and can be used as a basis
for the generation of entanglement by linear optics. While strong squeezing is challenging to generate in optical fields, here we present simple periodic modulation protocols in optomechanical systems that can generate large squeezing of their mechanical degrees of freedom for realistic system parameters. We then proceed to show how such protocols can serve to improve the measurement precision of weak forces and enhance the generation of entanglement between test masses that are subject to any kind of weak interaction. Moreover, these protocols can be reverted to reduce the amount of injected energy, while preserving the generated entanglement and making it more resilient to noise. We present the principle at work, discuss its application in a variety of physical settings, including levitated and tethered mechanical harmonic oscillators, and present example applications to Casimir and gravitational forces.

\end{abstract}

\maketitle
{\it Introduction.---}The quantum control of the mechanical degrees of freedom of ever more massive objects is one of the permanent ambitions of quantum physics. On the lower end of the mass spectrum, trapped ion technologies~\cite{LeibfriedBM+2003,Wineland+2013} and  atom interferometers \cite{CroninSP+2009} represent, arguably, the state of the art on the control of the mechanical properties of a particle. Above that level, the realm of physical systems for which exquisite control has been achieved extends across many orders of magnitude in mass, from the atomic to the mesoscopic scale. This includes macromolecules containing up to $10^4$ amu~\cite{EibenbergerGA+2013}, Bose-Einstein condensates of $\sim 10^9$ atoms~\cite{FriedKW+1998}, as well as solid state systems that can be either clamped resonators in a wide variety of implementations~\cite{AspelmeyerKM+2014}, e.g. pendula of remarkable stability~\cite{CatanoSE+2020}, or levitated nanoparticles~\cite{ChangRP+2010,MillenMP+2020} that promise extended coherence times due to their extraordinary isolation from the environment. Despite the many challenges that come with each mass regime and setup, the intense experimental activity of the last decade in the field of optomechanics has lead to the conquest of encouraging milestones. These include the ground state cooling of clamped~\cite{TeufelDL+2011, ChanMS+2011} as well as levitated resonators~\cite{DelicRD+2020}, or the remote entanglement between two tethered massive oscillators mediated by optical fields~\cite{RiedingerWM+2018, OckeloenDP+2018}, suggesting that extensive quantum control of mechanical properties well beyond the atomic scale might be within reach in the near future. This would have two main implications: on the one hand, it would open the door to experimentally testing properties of quantum mechanics such as its linearity at an unprecedented scale of mass and size, and with it allow to verify or falsify extensions of the Schr\"odinger equation like collapse models~\cite{BassiLS+2013}. On the other hand, attaining quantum control of the mechanical degrees of freedom of a massive resonator would entail applications in precision sensing of minute forces~\cite {HebestreitFR+2018,KuhnSK+2017,PedernalesMP+2020,PedernalesCP+2020}, and in particular of forces that are proportional to the size of the detector, like Casimir-Polder or gravitational forces \cite {SchneiterQS+2020,QvarfortPB+2020}.

In the light of the high force sensitivities that such devices augur, a natural question arises: would two such resonators be sensitive to the mutually induced weak forces that emerge when placed sufficiently close to each other? If the answer is positive, measuring the generated correlations would provide information on the nature of these forces ~\cite{Feynman+1957,KafriT+2013,KrisnadaKZ+2017} in possible experiments based on optomechanical and other technologies~\cite{PedernalesCP+2020,PedernalesMP+2020,LindnerP2005,KafriTM+2014,BahramiBM+2015,SchmoleDH+2016,BoseMM+2017, MarlettoV+2017,KrisnandaTP+2020,QvarfortBS+2020}. Not only that, provided that entanglement could be mediated by such weak forces, a complete toolbox of QED operations would become immediately available for massive optomechanical setups, with a plethora of applications in quantum information and metrology. However, the low energy of these interactions inevitably challenges this ambition.


\begin{figure}[!t]
\begin {center}
   \includegraphics[width=\columnwidth]{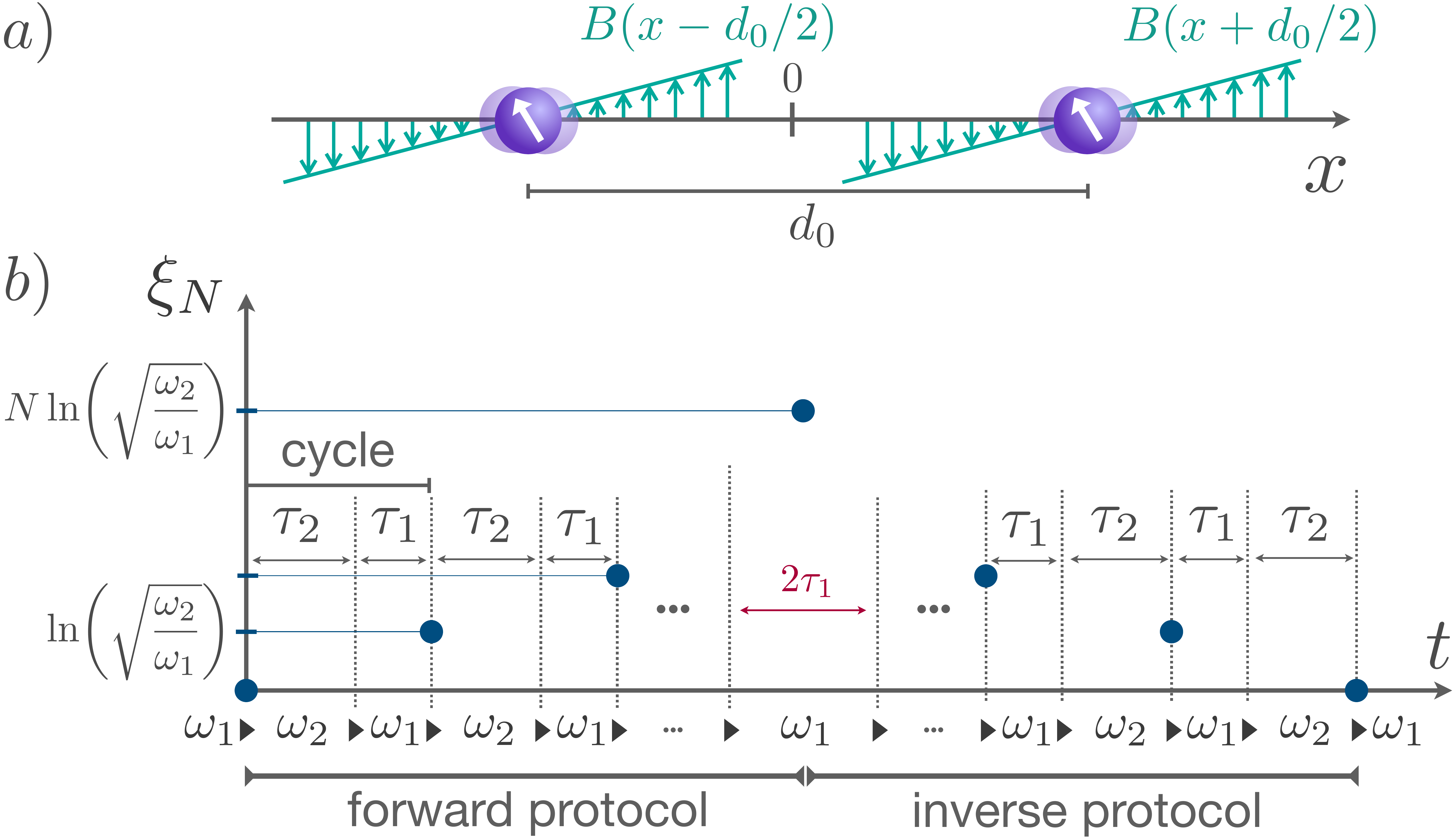} 
\end{center}
\caption {(a) Pictorial representation of a possible setup. Two diamagnetic nanoparticles 
potentially hosting a spin degree of freedom, are trapped in two linear magnetic field gradients. The particles oscillate around equilibrium positions determined by the zeros of the magnetic fields, which are separated by a distance $d_0$, and interact via a weak force. (b) Squeezing protocol. Each particle is subject to a sequence of jumps between two frequencies $\omega_1$ and $\omega_2$ properly spaced in time by intervals $\tau_1$ and $\tau_2$, respectively. After N cycles, this leads to squeezing of the mechanical degrees of freedom of each oscillator, parametrized by a squeezing parameter $\xi_N$ that grows linearly in time. The squeezing can be reverted by applying the inverse protocol.}
\label {fig:sketch}
\end{figure}


In this Letter, we show that two weakly coupled nanomechanical resonators can develop sizeable entanglement between their mechanical degrees of freedom even in the presence of dissipation, provided that these are subject to a continuously applied protocol that squeezes them in suitable quadratures, and that this squeezing is removed sufficiently fast  after the desired entanglement has been generated. We describe analytically and simulate numerically such a protocol, consisting of periodically applied local shifts of the resonator frequencies. We show how our protocol can generate squeezing in the system to accelerate the generation of entanglement, and subsequently reduce the squeezing again, while retaining the generated quantum correlations, in order to avoid a deleterious enhancement of the sensitivity to environmental noise. This extraction of local squeezing also facilitates the verification of entanglement, which can be carried out quantitatively by the detection of correlations in local measurements ~\cite{audenaert2006correlations,eisert2007quantitative,guhne2007estimating}.

{\it Setup.---}The central findings of our analysis are applicable to any setup consisting of two weakly interacting massive resonators operating at low temperatures. This includes nanoparticles levitated, either by optical means~\cite{DelicRD+2020}, in Paul traps~\cite{DelordNB+2017}, or diamagnetically~\cite{HsuJL+2016}, as well as tethered oscillators such as massive pendula \cite{MatsumotoCM+2019}. The origin of the weak interaction can be, among others, a surface force like the Casimir-Polder force or a gravitational force. For the sake of simplicity, we consider two identical oscillators, that is, with the same mass $m$ and trap frequency $\omega$, and consider the motion in only one spatial direction, e.g. the $x$ direction. Assuming that the centers of the harmonic traps confining the two oscillators are separated by some distance $d_0$, see Fig.~\ref{fig:sketch} (a), the  Hamiltonian of the system is given by    
\begin {equation}
\begin {split}
 H_{\mathrm D}=U_{\mathrm W} (x_1,x_2)+\sum_{i=1}^2 \frac { p_i^2} {2 m} + \frac {1} {2} m \omega^2 ( x_i+ d_i)^2,
\label {d-ham}
\end {split}
\end {equation}
where $ U_{\mathrm W}$ represents the weak coupling between the resonators and $d_{1/2}= \pm d_0/2$. For the sake of generality, we consider that the interaction energy is a function of the  separation distance between the centers of mass of the two resonators, $d=\abs{x_1 - x_2}$, and follows an inverse power law with exponent $n$, $U_{\rm W} = C/\abs{d}^n$, such that we can expand it around the point $d = d_0$, provided that $\abs{d - d_0} \ll d_0$, as
\begin{equation}
    U_{\rm W} = \frac{C}{d_0^n}[(1 - \frac{n}{d_0}(d - d_0) + \frac{n(n+1)}{2d_0^2}(d - d_0)^2 + \dots].
    \label{WeakInt}
\end{equation}
A coupling between the mechanical degrees of freedom of the two nanoparticles is established by the third term in the expansion. In this quadratic form the interaction Hamiltonian is very general and allows to replicate in this platform well established protocols from quantum information processing platforms like trapped ions~\cite{SerafiniRP+2009,SerafiniRP+2009b}. Our purpose is to estimate the time scale required to generate detectable entanglement mediated by such a weak interaction and check if it is compatible with the state of the art. This would allow us to entangle two massive oscillators without the use of charges or magnetic impurities, which are potential sources of noise, and could help us determine experimentally the {\it{`quantumness'}} of these forces.

For the specific parameters  in Eq.~(\ref{d-ham}), when running numeric simulations, we will consider values corresponding to magnetically levitated diamagnetic particles. These have promising prospects in terms of coherence when compared to tethered oscillators, as the detachment from the substrate dramatically reduces their interaction with the environment, and also when compared to optically levitated nanoparticles, as the passive trapping fields remove the dissipation mechanisms associated with the recoil and absorption of photons from the optical trapping field. Stable levitation of diamagnetic nanoparticles has been demonstrated employing magnetic field gradients from permanent magnets in high vacuum, with center of mass motion temperatures below 1 mK~\cite{HsuJL+2016,SlezakLH+2018}. Such setups represent a promising platform to implement matter-wave interferometry at macroscopic scales, for high-sensitivity quantum metrology, and for investigating the fundamental limits of quantum mechanics~\cite{BassiLS+2013, Romero+2011,PinoPS+2018}.

For a diamagnetic particle  in a magnetic field $B(x)$, the  potential energy is given by $U_B= - \frac {m \chi B(x)^2}{2 \mu_0 \rho}$, where $m$ is the mass of the particle, $\rho$ its density, $\mu_0$ the vacuum permeability, and $\chi$ is the material magnetic susceptibility. In case of a linear magnetic gradient such as $B(x)= B' x$, and negative magnetic susceptibility (diamagnetism), the magnetic energy acts as an effective harmonic potential with frequency $\omega  = \sqrt{ -\frac {\chi}{\mu_0 \rho}}B'$, proportional to the magnetic field gradient $B'$. For diamonds the magnetic susceptibility is $\chi = -2.1 \times 10^{-5}$ and  the density $\rho =3500\,\mathrm{Kg/m}^3$, allowing trapping frequencies of the order of $\sim 2\pi\ 100$~Hz, for magnetic field gradients of the order of $\sim 10^4$~T/m. We stress that considering different materials leaves the principal conclusions of this analysis unchanged.

As a particular instance of the weak interaction in Eq.~(\ref{WeakInt}), we will consider the Casimir-Polder force~\cite{GarrettSM+2018,ChumakMB+2004}, which arises between any two surfaces in proximity, as a consequence of the quantization of the electromagnetic field. Originally derived for two parallel metallic plates, it can also be computed for compact objects of arbitrary shape and material~\cite {EmigGJ+2007}. For two identical spheres, the leading contribution to the interaction energy takes the form~\cite{EmigGJ+2007} ${U_{\mathrm C} (x_1,x_2) =\alpha {R_0^6/\abs{x_1-  x_2}^7}}$, where $\alpha$ is a function of the electric and magnetic permeabilities of the spheres~\cite{footnote2}.

{\it Entanglement.---}In order to highlight the challenges involved, let us consider two oscillators of frequency $\omega$ prepared in their ground states. Under a coupling term of the form of Eq.~\eqref {WeakInt}, with $C=\alpha R_0^6$ and $n=7$, corresponding to a Casimir interaction, the system will develop entanglement between its mechanical degrees of freedom, which quantified by the logarithmic negativity~\cite{Plenio2005,AudenaertEP+2008}, oscillates in time reaching a maximum value of $E_N^{\mathrm {max}}\simeq 56 \alpha R_0^6/(m \omega^2 d_0^9 \ln{2}) $. For two  diamonds with a radius of $250$~nm, which are held at a distance $d_0 =5 \, \mu$m and trapped with a frequency $\omega/2\pi= 100$ Hz, we estimate a maximum logarithmic negativity $E_N^{\mathrm {max}}\simeq 10^{-6}$. However, this macroscopic manifestation of quantum effects is very fragile and is dramatically reduced when a realistic dissipative dynamics of the oscillators is considered. This sets daunting perspectives in the feasibility of detecting such an entanglement unless suitable countermeasures are taken.


\begin{figure}[hbt!]
\begin {center}
   \includegraphics[width=\columnwidth]{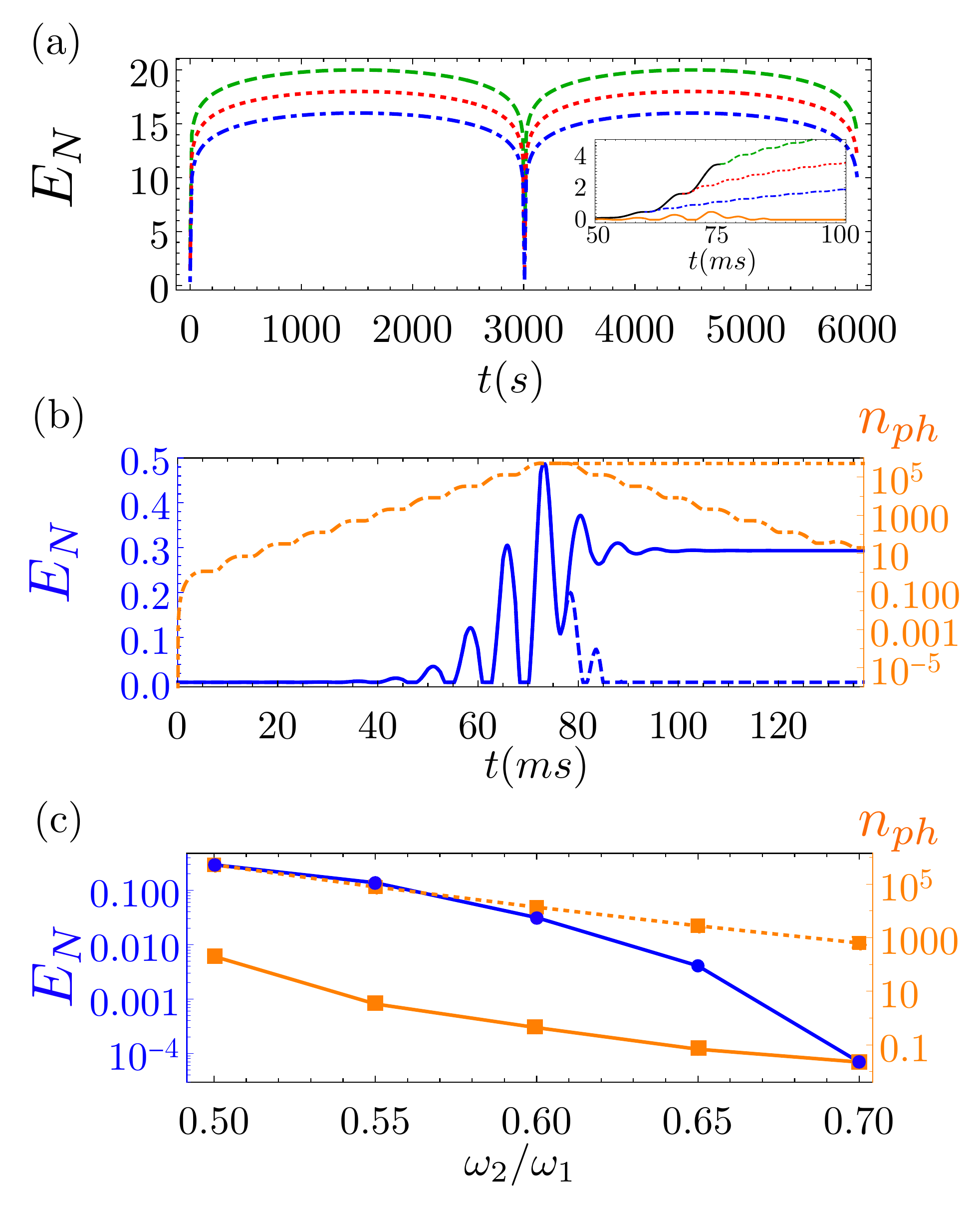} 
\end{center}
\caption{Time evolution of the logarithmic negativity between two trapped nanodiamonds of radius $R_0=250$~nm, held at a distance $d_0 =5\, \mu$m and coupled through Casimir interaction. The system is initially in the ground state and undergoes a sequence of periodic frequency shifts between $\omega_1=2\pi\  100$~Hz and $\omega_2=0.5 \omega_1$. (a) Long-time dynamics of the logarithmic negativity for different durations of the squeezing protocol, without dissipation. In dot-dashed blue, dotted red, and dashed green, we show the entanglement generated after squeezing trough 8, 9, and 10 cycles, respectively. In the inset, the short-time dynamics, displaying in black (upper solid line) the evolution during the squeezing protocol and in orange (lower solid line) the case in which dissipation is considered for a protocol with 10 cycles. (b) Dynamics of entanglement for a protocol consisting of $N=10$ cycles forward and another 10 cycles backward (solid blue line) and for the case with no backward cycles (dashed blue line). In orange the number of phonons in the oscillators during the protocols,  dot-dashed orange  for the complete protocol, and dotted orange for the evolution without the reversal. (c) Logarithmic negativity at the end of a protocol as in (b),  as a function of the ratio between the two alternating frequencies (blue circle markers). In orange (square markers),  the number of phonons before the reversal (dashed) and at the end of the protocol (solid). In (b) and (c), we have included the decoherence effects induced by the interaction with a thermal bath with $\bar n= 100$ phonons and coupled to the system with a rate $\Gamma=\omega_1/Q$, for a quality factor of $Q=10^8$ \cite {supp-mat}.}
\label{fig:entanglement}
\end{figure}


Thereby, here we resort to strategies that can increase the effective interaction between the two subsystems. An established way to enhance the sensitivity of a system to external interactions consists in squeezing its mechanical degrees of freedom. For each oscillator, one method to achieve this is to suddenly relax the frequency of the trap from $\omega_1$ to a lower value $\omega_2$, wait for a quarter of a period of the new trap frequency, $\tau_2= \frac{\pi}{2 \omega_2}$, then switch the frequency back to the initial value $\omega_1$ and wait for another quarter of a period of the current trap frequency, $\tau_1= \frac{\pi}{2 \omega_1}$. This results in the squeezing of the initial state of the oscillator $\rho_{\rm in}$ by an amount $\xi= \ln (\sqrt{\omega_2/\omega_1})$, where the new state is $S(\xi) \rho_{\rm in} S(\xi)^\dag$, with $S(\xi)$ the squeezing operator \cite {RashidTB+2016}. However, in experiments, the range of available frequencies is, typically, limited. In particular, for magnetic traps, this limitation is imposed by the attainable magnetic field gradients, which have maximum values, usually, on the order of $\sim 10^4$ T/m. Nonetheless, it is possible to enhance the generation of squeezing beyond $\xi$ by periodically repeating such a frequency jump protocol. As a matter of fact, by repeatedly switching between two frequencies, $\omega_1$ and $\omega_2$, the amount of squeezing can be made to increase linearly in time, provided that frequency jumps are properly timed~\cite {JanszkyY+1986, MaR+1989, JanskyA+1992, KissAJ+1994}, see Fig.~\ref{fig:sketch}(b). After $N$ cycles, each one lasting $\tau_1+\tau_2$, the resulting squeezing parameter is  $\xi_N=N \ln (\sqrt{\omega_2/\omega_1})$. With this proposed scheme, a restricted choice of experimentally accessible frequencies can still allow for the generation of a large amount of squeezing.

In the presence of interaction, local squeezing enhances the effective coupling as $S(\xi)^\dag x_1 x_2 S(\xi) = e^{2r} x_1x_2$, with $\xi = -r$ and $r$ real \cite{LauC+2019,BondarBC+2020}. A variety of protocols relying on squeezing have been developed and proposed as a standard scheme to enhance the generation of entanglement in different platforms, from charged particles as trapped ions~\cite{SerafiniRP+2009,SerafiniRP+2009b} to neutral objects as massive bodies interacting via gravitational interaction~\cite {KrisnandaTP+2020}. In Fig.~\ref{fig:entanglement} (a), we show the evolution of entanglement for two nanodiamonds interacting via Casimir-Polder forces, which are initially subject, for different periods of time, to the frequency jump protocol described above. We observe that entanglement oscillates with a period determined by the strength of the interaction and an amplitude determined by the amount of injected squeezing \cite {YeomanB+1993,supp-mat}, which in turn is proportional to the duration of the protocol. In the inset, the orange line shows the evolution of the entanglement for the same system, but now in the presence of dissipative noise of the oscillators. We observe the strong deleterious effect that even an optimistic dissipation rate has on the entanglement. While, during the action of the protocol, entanglement is still generated in the presence of sufficiently weak dissipation, albeit at a much slower rate, it quickly decays as soon as the protocol is stopped. 

As the decay rate of the entanglement is expected to be proportional to the generated squeezing, we extend our protocol to contain an inverse sequence of frequency jumps, which is able to reduce the local squeezing that has built up during the protocol, and with it reduce the environmental sensitivity of the system. Despite the continued presence of the mutual interaction, this protocol is found not to affect the entanglement that has built up.  This is one of the main findings of our analysis, and shows that the complete sequence of pulses constitutes a valid sensing protocol able to the selectively enhance the signal over the noise. Besides, the reduction of the system energy has the additional beneficial side effect that it tends to facilitate the measurement of observables at the end of the protocol, as the number of energy levels that requires control is reduced.  More specifically, the reversal is achieved by running the inverse sequence of frequency jumps after a waiting time of half of a period of the trap frequency, see Fig.~\ref{fig:sketch} (b). In the absence of noise, control imperfections, and of interaction between the particles, after running such an inverse sequence, all the injected squeezing would be extracted, and the system would return to its initial state. For a single particle, this would constitute a witness of coherence over the spatial length scale of the maximally squeezed state reached at the middle of the protocol, which could be used to set bounds on collapse models~\cite{BassiLS+2013, supp-mat}. In the presence of interaction, however, the system will retain the acquired entanglement, and will, consequently, not return to its initial state. Nevertheless, the amount of excitations gained over the complete protocol can be regarded as the smoking gun for the presence of the weak interaction. 

In Fig.~\ref{fig:entanglement} (b), we plot the time evolution of the logarithmic negativity for the same setup discussed in (a), but now applying the reverse protocol. The continuous blue line shows how the inversion of the protocol is able to dramatically reduce the decay rate of the entanglement, as compared to the case where no inversion protocol is applied, dashed blue line. The orange dot-dashed and dotted lines show the evolution of the excitations in the resonators for the protocol with and without the inversion part, respectively. It is verified that the inversion of the protocol strongly reduces the energy of the system, albeit a number of excitations is retained in the system as a consequence of the generated entanglement. The precise energy reduction achieved by the the reverse protocol is provided in Fig.~(\ref{fig:entanglement}c) as a function of the ratio of frequencies employed in the protocol. In blue circles, the entanglement achieved at the end of the protocol for each case is displayed \cite{footnote3}. 

On the other hand, recent developments in tethered pendula~\cite{MatsumotoCM+2019,CatanoSE+2020} with remarkable stabilities, encourage us to consider similar protocols in such setups, now, with gravitationally mediated interactions. This is motivated by the idea that the {\it `quantumness'} of a mediating force can be assessed by its ability to act as a quantum channel~\cite {Feynman+1957,KafriT+2013,KrisnadaKZ+2017}. Consider the pendulum in Ref.~\cite{MatsumotoCM+2019}, with an eigenfrequency of $\omega_1 = \sqrt{g/l} = (2\pi)\ 2.2$ Hz. A frequency shift can, for example, be induced by suddenly pulling the base of the pendulum upwards with an acceleration $a_{\rm up}$. Ignoring finite material stiffness, this would result in a new oscillation frequency $\omega_2 =\sqrt{(g+a_{\rm up})/l}$. For example, an acceleration of $a_{\rm up} = g$ would result in a frequency shift $\omega_2/\omega_1 = \sqrt{2}$, while the base of the pendulum would be displaced by an amount $g\tau_2^2/2 = (\pi/4)^2l$, after a quarter of the new period. We estimate that two such pendula, which have a mass of $m = 7$ mg, placed at a distance of $2$ mm from each other, and interacting only gravitationally, would develop in a time of $t= 10$ s an entanglement of $E_N\simeq 0.5$, in the presence of a decoherence process characterized by $\bar n/Q = 10^{-10}$, where $\bar n$ is the average phonon number of the thermal bath, and $Q$ the quality factor of the pendula. This confirms the extremely challenging isolation conditions required to observe gravitationally mediated entanglement between massive resonators, which, as expected, remains far more demanding than the earlier discussed Casimir mediated case.

Once a significant amount of entanglement has developed in the system, the question 
arises on how to detect it. As the full state may not be Gaussian due to small
experimental imperfections as well as higher order contributions in Eq.~\ref{WeakInt}, 
one needs to adopt the approach developed in~\cite{audenaert2006correlations} and 
determine the least entangled state, quantified by the logarithmic negativity, that
is compatible with, for example, the measured covariance matrix. An alternative approach is available when the oscillators can be coupled to spin degrees of freedom~\cite{KolkowitzJU+2012, GieselerKR2020}. This can be the case, for example, if one considers the oscillators to be nanoparticles hosting a color center, e. g. NV centers in diamond~\cite{NeukirchGQ+2013,PettitNZ+2017, GeiselmannJR+2013}. In such a setup, the spins can be coupled to the motion of the oscillators by placing a spatially inhomogeneous magnetic field. This interaction can then be exploited, either to map the already generated entanglement onto the spins, or to make the spins interact through the weak force between the oscillators. In the second case, an enhancement of the interaction can again be achieved by squeezing the mechanical resonators with the protocol introduced here and combined with other techniques designed to increase noise resilience \cite {ViolaKL+1999,MuelleKC+2014,CasanovaHW+2015,ArrazolaCP+2018}. The specific details of such a protocol are outside of the scope of this work and will be described in coming publications.

{\it Conclusions.---} Optomechanics is ushering in a new era of quantum resonators with unprecedented quality factors that is opening the door to quantum optical experiments in new mass regimes. We believe that it is possible to extend these platforms to accommodate more than one massive oscillator that interact with each other via weak forces. We have presented a protocol that can both, inject and extract large amounts of local squeezing in the resonators mechanical degrees of freedom with modest resources, by periodically modulating their frequencies. This combination of squeezing injection and extraction allows to amplify the sensitivity of the system to weak interactions for a short time window and then attenuate it again, in order to recover resilience to noise without loosing the acquired entanglement. This establishes an interaction channel between massive nanomechanical resonators that is not destroyed by its weakness or noise sensitivity, and which, if quantum in nature, it can be used to generate quantum correlations between the spatially separated particles, thus providing means to examine the very nature of these interactions~\cite{PedernalesCP+2020,PedernalesMP+2020,Feynman+1957,LindnerP2005,KafriT+2013,KafriTM+2014,BahramiBM+2015,SchmoleDH+2016,KrisnadaKZ+2017,BoseMM+2017,MarlettoV+2017,KrisnandaTP+2020,QvarfortBS+2020}. Furthermore, this offers a potential interface between optomechanics and existing quantum platforms which can be used to import many quantum control techniques that have already proven successful. We believe that our findings set the ground for a multi-particle quantum platform operating in a new mass regime.
\begin{acknowledgments}
{\it Acknowledgments.---} We acknowledge support by the ERC Synergy grant HyperQ (Grant No. 856432), the EU projects HYPERDIAMOND (Grant No. 667192) and AsteriQs (Grant No. 820394), the QuantERA project NanoSpin, the BMBF project DiaPol, the state of Baden-W\"urttemberg through bwHPC, the German Research Foundation (DFG) through Grant No. INST 40/467-1 FUGG, and the Alexander von Humboldt Foundation through a postdoctoral fellowship.

Note added. While preparing this work for journal submission we became aware of \cite{Weiss+20} which pursues closely related ideas.
\end{acknowledgments}
\bibliographystyle{apsrev4-1}

\begin{thebibliography}{99}
\bibitem {footnote3} For every simulation, we check that $\sqrt{\langle (x_1-x_2)^2 \rangle} \ll d_0$ at all times, such that the harmonic approximation employed for the Casimir interaction is always valid. For a thermal initial state with $n_{ph}$ phonons, this translates to the condition $ (\frac {\omega_1}{\omega_2})^N\sqrt{\frac {\hbar}{2 m \omega_1} (2 n_{ph}+1)} \ll d_0$, where we assume that  $\omega_1 > \omega_2$.
\end{thebibliography}

\clearpage
\appendix

\section{Entanglement under xx interaction}
\label{appendix:xxinteraction}
An interaction of the form $H_{\rm int} = \lambda x_1x_2$, as considered in the main text, can be divided into two terms: the beam splitter term, $H_{\rm bs}= \lambda x_0^2 (a^\dag b + ab^\dag)$, and the two-mode squeezing term, $H_{\rm tms} = \lambda x_0^2(ab + a^\dag b^\dag)$, where $\{a,a^\dag\}$ and $\{b,b^\dag\}$ are ladder operators for each of the two modes, such that $x_1 = x_0(a + a^\dag)$ and $x_2 = x_0(b + b^\dag)$. While both forms of interaction have the ability to entangle an initially separable state, if considered separately, the dynamics of the generated entanglement is rather different. For a time evolution under the beam splitter type of interaction, the amount of created entanglement will depend on the initial state, for example, if applied to vacuum such an interaction will generate no entanglement, while, when applied on two squeezed states, these will effectively evolve, at short times, as under a two-mode squeezing operation acting on the vacuum~\cite{YeomanB+1993}. Nevertheless, for any initial state, the maximum amount of entanglement is bounded in time, and will oscillate with a period determined by the coupling strength, $T = 2\pi/(\lambda x_0^2)$. On the other hand, an interaction of the type $H_{\rm tms}$ will entangle the two modes irrespective of the presence of squeezing in the initial state and, furthermore, this entanglement is not bounded in time, that is, it will grow linearly instead of oscillate. Moreover, numerical simulations suggest that for the same amount of entanglement a state generated by applying the two-mode squeezing interaction to the vacuum contains less energy and is more resilient to dissipation than a state generated by inputting two squeezed states to the beam splitter type of interaction. This is not all that surprising, as the latter contains single-mode squeezing, which makes it more sensitive to dissipation.

For the $xx$-type of interaction, in the presence of trapping potentials of similar frequency for each of the two modes, the beam splitter component of the interaction will be dominant, while the two-mode squeezing component will be off resonant. This explains the oscillations observed in the long-time dynamics of the logarithmic negativity, in Fig.~(2a) of the main text. However, it would be desirable to have the opposite situation, that is, the two-mode squeezing part of the interaction to be resonant, while the beam splitter part is off resonant, as this form of interaction leads to a constant growth of the entanglement, reaching higher values and being more resilient to dissipation. One natural way to reach this situation is to invert the frequency of one of the resonators $\omega \rightarrow -\omega$, however, such an operation may not be available in many setups. An alternative approach is to mimic dynamical decoupling techniques for spins~\cite{ViolaKL+1999} where, by flipping the sign of spin operators in resonance with their oscillation frequency, one is able to bring specific Hamiltonian terms in and out of resonance, either for noise resilience or to amplify specific interactions~\cite{MuelleKC+2014, CasanovaHW+2015, ArrazolaCP+2018}. In spin systems, the way to change the sign of a given spin operator is to apply a fast $\pi$-pulse in a direction that is orthogonal to the operator. For bosonic modes, the equivalent effect can be achieved by suddenly changing the frequency of the oscillator to a value that is much larger, such that $\omega \rightarrow \tilde \omega$ with $\tilde \omega \gg \omega$, and after waiting for half of a period of the new frequency, $\tau = \frac{\pi}{\tilde \omega}$, switching back to the original frequency. This protocol would have the effect of mapping $\{ b, b^\dag \} \rightarrow - \{ b, b^\dag \}$, in a time scale that can be regarded as instantaneous for the frequency of the modes, $\tau \ll 1/\omega$. Using this technique to flip the sign of one of the two resonators with the periodicity of $\omega$ one would achieve a resonant two-mode squeezing type of interaction. We relegate the analysis of the performance of such a protocol under realistic experimental conditions to future work.

\section{Numerical simulations}
\label{appendix:numerics}

To compute the time evolution of the interacting resonators, we use the covariance matrix formalism, as we always consider initial Gaussian states, and the Hamiltonian is quadratic. We consider a system of interacting resonators evolving under a Hamiltonian of the form
\begin{equation}
H = \frac {1}{2} \va{r}^T \left( \begin{array}{cccc}
    m \omega^2 & \lambda & 0 & 0 \\
    \lambda & m \omega^2 & 0 & 0 \\
    0 & 0 & \frac {1}{m} & 0 \\
    0 & 0 & 0 & \frac {1} {m}
\end{array} \right) \va{r},
\end{equation}
with $\va{r}(t)= \left(\begin{array}{cccc}
           \hat{x}_1 & \hat{x}_2 & \hat{p}_1 & \hat{p}_2
          \end{array}\right)^T$, and define the covariance matrix of a quantum state $\rho$ as
\begin{equation}
V_{ij}(t) =\frac{1}{2}\expval{r_i(t)r_j(t)+r_j(t)r_i(t)}-\expval{r_i(t)}\expval{r_j(t)}.
\label {free-ham}
\end{equation}
It can be shown that the covariance matrix obeys a Markovian master equation of the form~\cite{SerafiniRP+2009}
\begin {equation}
\frac {d}{dt} V= K V+ V K^T-\Gamma V+ \Gamma V_{\infty},
\label{cv-meq}
\end {equation}
where $\Gamma$ is a Markovian decay rate, used to define the quality factor as $Q = \omega/\Gamma$, and $V_{\infty}$ is the covariance matrix of the system in thermal equilibrium with the environment, that is, the tensor product of two thermal states of resonators of frequencies $\omega_1$ and $\omega_2$ at a given temperature $T$. The matrix $K$ describes the unitary evolution induced by \eqref {free-ham} and has the form
\begin{equation}
K = \left( \begin{array}{cccc}
0 & 0 & \frac {1} {m} & 0 \\
0 & 0 & 0 & \frac {1} {m} \\
-m \omega^2 & - \lambda & 0 & 0 \\
-\lambda & -m \omega^2 & 0 & 0
\end{array} \right).
\end{equation}
The formal solution of \eqref {cv-meq} is found to be
\begin{equation}
\label{eqmotcov}
\begin{split}
V(t) = e^{K(t-t_0)} V(t_0)e^{K^T(t-t_0)}e^{-\Gamma (t-t_0)} \\
+\Gamma \int_{t_0}^t dt' e^{-\Gamma (t-t')} e^{-K(t-t')} V_{\infty} e^{-K^T(t-t')}.
\end{split}
\end{equation}

For a fixed set of Hamiltonian parameters, the integral in Eq.~(\ref{eqmotcov}) can be analytically solved. For the data reported in the manuscript, we numerically solve the equations of motion including the periodic switch between two values of the oscillator frequencies. Once obtained the covariance matrix, the entanglement at a given time t is computed from it in the form of the logarithmic negativity. In general, for a system of $m$ modes in the Gaussian quantum state $\rho$, equivalently characterised by the covariance matrix $V$, the logarithmic negativity quantifying the entanglement between two complementary subsystems of the modes is given by
\begin{equation}
    E_N(\rho) = - \sum_{k=1}^{2m} \log_2[\min(1,2\abs{\nu_k})],
\end{equation}
where $\nu_k$ are the symplectic eigenvalues of the covariance matrix $V^{\Gamma}$ of 
the partially transposed states $\rho^{\Gamma}$, that is, the eigenvalues of the matrix 
$i \Omega V^{\Gamma}$, where $\Omega$ is the symplectic matrix 
\begin{equation}
    \Omega = \left( \begin{array}{cc} 0 & I_m \\ - I_m & 0 \end{array} \right),
\end{equation}
with $I_m$ the identity matrix of dimension $m$, and the superscript $\Gamma$ represents the  partial transposition with respect to one of the two subsystems~\cite{AudenaertEP+2008}. 
The covariance matrix of the partially transposed density operator $\rho^{\Gamma}$ 
with respect to the the first $q$ modes is given by $V^{\Gamma}=PVP$ where $P=I_m\oplus I_q \oplus (-I_{m-q})$. For the case of only two modes, this expression can be reduced to 
\begin{equation}
 E_N(\rho)=  - 2\log_2[\min(1,2|\nu_{\rm min}|)],
\end{equation}
where $\nu_{\rm min}$ is now the minimum symplectic eigenvalue of the covariance matrix after partial transposition w.r.t. one of the two modes.


\begin{figure}[!h]
\begin {center}
   \includegraphics[width=\columnwidth]{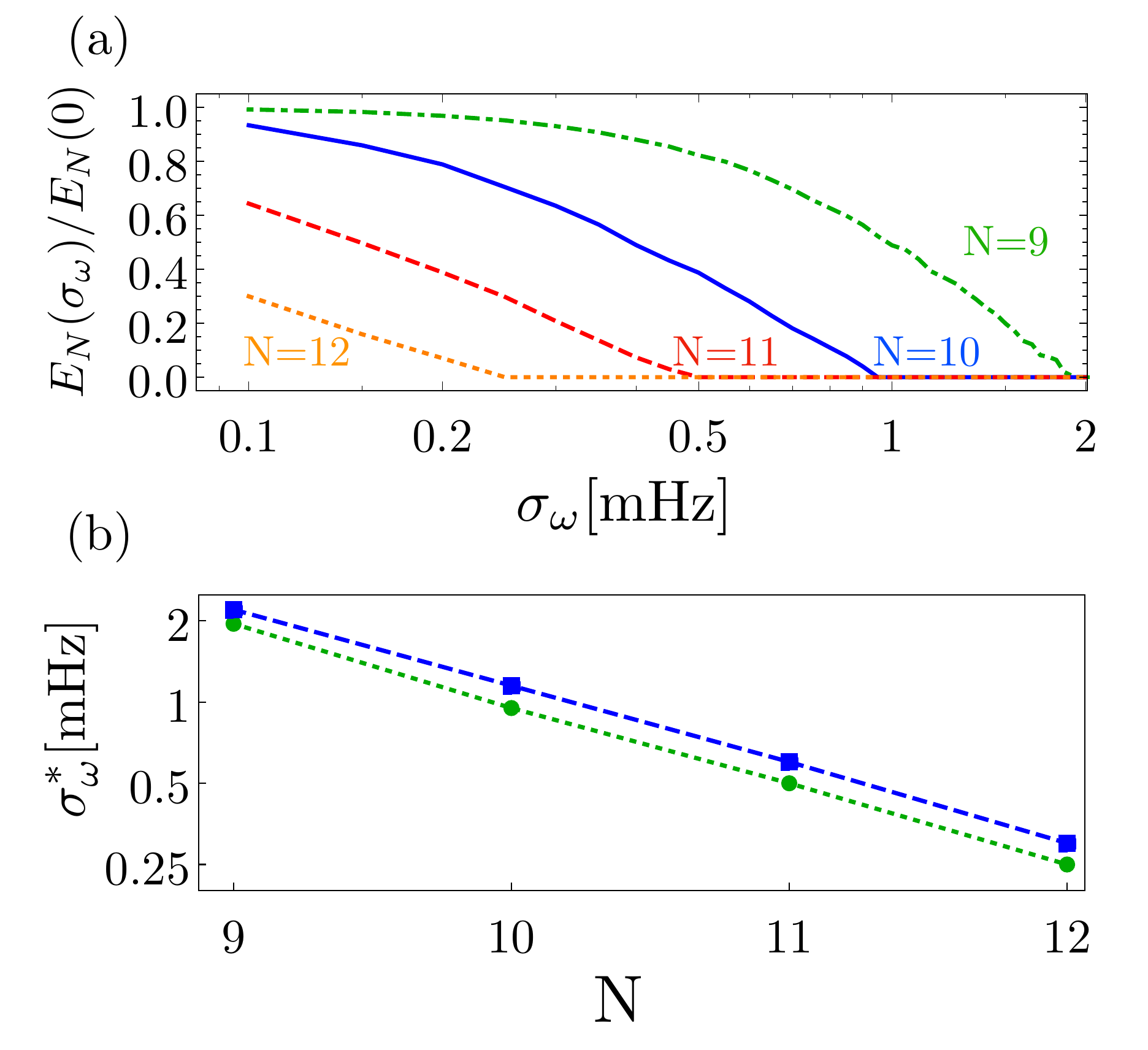} 
\end{center}
\caption {A system of two particles of radius $R_0=250$~nm, with equilibrium positions separated by $d_0 =5\, \mu$m and coupled through Casimir interaction, is prepared in the ground state of its uncoupled Hamiltonian and periodically undergoes a sequence of imperfect frequency shifts between the ideal values $\omega_1=(2\pi)\  100$~Hz and $\omega_2=0.5 \omega_1$. (a) Normalized logarithmic negativity at the end of a protocol consisting of $N$ cycles  forward  and  another  $N$  cycles  backward as a function of the variance in frequency. The normalization is with respect to the ideal case (zero variance) for each protocol. (b) Values for the variance above which no entanglement is observed, $\sigma_\omega^*$, as a function of the  number of cycles $N$ in the protocol. In blue we display $\sigma_\omega^*$ for the forward protocol only, and in green we display $\sigma_\omega^*$ at the end of the protocol. }
\label {fig:sup}
\end{figure}


\section{Sensitivity due to control errors}
In the lab, a frequency-jump protocol as the one discussed in the main text will be subject to a finite precision in the ability to set a specific frequency in each step of the protocol. In this section, we analyse the sensitivity of our protocol to control errors on the performed frequency jumps, and bound the precision required to observe entanglement. For that, we assume that after each frequency jump in the protocol the attained frequency deviates from the ideal one by a random amount with a finite variance $\sigma_\omega$, such that the implemented frequency is randomly picked from a Gaussian probability distribution centered around the ideal value, $p(\omega)= \frac {1}{\sqrt{2 \pi \sigma_\omega^2}}\exp (-\frac{(\omega-\omega_{1/2})^2}{2 \sigma_\omega^2})$. We then calculate numerically the logarithmic negativity obtained from the covariance matrix after averaging it over this source of noise. Here, we make the simplifying assumption that the averaged state is still Gaussian and, therefore, fully characterized by the averaged covariance matrix. The results of our simulations are displayed in panel (a) of Fig.~\ref{fig:sup}, where we show the logarithmic negativity  after a complete protocol (consisting of the forward and backward sequence of pulses) as a function of the variance in the frequencies. As expected, we observe that an uncertainty on the frequencies employed in the sequence reduces the amount of entanglement retained at the end of the protocol. Specifically, in the regime of parameters considered here (see the caption of  Fig.~\ref{fig:sup} for details), we notice that a precision below mHz is required in order to observe entanglement. Furthermore, we see that the tolerable noise depends on the length of the protocol. In panel (b) of Fig.~\ref{fig:sup} we display $\sigma_\omega^*$, defined as the threshold value of the variance for which the entanglement is completely destroyed. We see how this threshold values decrease with the number of cycles. This behaviour could be qualitatively understood as follows: a frequency shifted by $\Delta \omega$, acting on the system for a time $t$, rotates the state of the system in phase space by an extra angle $\Delta \omega t$. Due to this rotation, after a quarter of a period of the nominal frequency, the position variance will change by an amount proportional to the momentum variance, that is, by an amount $\sim \frac {\Delta p^2}{m^2 \omega_1^2}e^{2 r} \sin(\frac{\pi \Delta \omega  \pi}{4 \omega_1}) $. In order to observe entanglement, this fluctuation must be smaller than the uncertainty of the vacuum state. This implies that, once the average is taken into account, the variance on the frequency should satisfy (qualitatively) the condition $\Delta \omega \ll \frac {4 \omega_1}{\pi} e^{-2 r}$, where $r=N \log (\omega_1/\omega_2)$. We numerically verify that this condition predicts the correct order of magnitude. 

The requirements for the case of gravitational interaction between pendula as reported in Ref.~\cite{MatsumotoCM+2019,CatanoSE+2020} with $\omega_1=(2\pi) 2.2$~Hz and a mass of $m=7$~mg as discussed in the main text are even more stringent.


\section{Bounds on collapse models}
When considering a single particle, and in the absence of noise and control imperfections, after running both forward and backward sequences of our protocol, all the generated squeezing is extracted, and the system  returns to its initial state. The recovery of the initial state at the end of our protocol can be regarded as a witness of coherence over the spatial length scale of the maximally squeezed state, reached at the middle of the protocol. A certification of such a coherence could then be used to set bounds on collapse models or other decoherence mechanisms in general. As a figure of merit for this coherence, we resort to the purity of the quantum state at the end of the protocol, which we can easily compute from the covariance matrix as
\begin {equation}
\mathrm {P}= \frac {1}{\sqrt{\det V}}.
\label {purity}
\end {equation}
For a massive particle in a spatial superposition, collapse models, such as the mechanism of continuous spontaneous localisation (CSL), predict a coherence time $\tau_{\rm CSL} = \frac {1} {\Lambda_{\mathrm {CSL}} d^2}$, which depends on the mass of the particle and the size of the coherent superposition~\cite {Romero-Isart+2011}. Here, $d$ is the size of the spreading and $\Lambda_{\mathrm {CSL}}$ a coefficient depending on the number of nucleons and the free parameters of the CSL  model
\begin {equation}
\Lambda_{\mathrm {CSL}} = \frac {m^2}{m_0^2} \frac {\gamma^0_{\mathrm {CSL}}}{4 a_{\mathrm {CSL}}} f(R/a_{\mathrm {CSL}}),
\end {equation}
where $m$ is the mass of the body, $m_0$  the mass of a single nucleon, in the following example taken to be the mass of a carbon atom,  and $f(x)= \frac {6 }{x^2}\left [ 1- \frac {2}{x^2} + \left (1+  \frac {2 }{x^2} \right)e^{-x^2}\right ]$ \cite {CollettP+2003}. $\gamma^0_{\mathrm {CSL}}$ and $a_{\mathrm {CSL}}$ are the free parameters of the model that need to be experimentally determined, and which are conventionally predicted to take values on the order of $\gamma^0_{\mathrm {CSL}}=10^{-16}$~Hz and $a_{\mathrm {CSL}}=100$~nm~\cite{Romero-Isart+2011,BassiLS+2013}. 

In our scheme, the system reaches its maximum spatial spread, $\sigma_{\rm max}$, at the middle of the protocol. However, at this point the system is rotating in phase space at frequency $\omega_1$ and, therefore, the system spends there only an infinitesimal amount of time. To set bounds on the free parameters of collapse models, both the spatial spread of the superposition and its duration in time are relevant. To that end, we determine the amount of time that the system spends in a state with a position variance that is $\sigma \geq \sigma_{\max}/2$, $\tau = 2 \pi/(3 \omega_1) = 4/3 \tau_1$, and we require that the collapse model allows for a coherence time that is at least one order of magnitude larger than that, $\tau_{\rm CSL} > 10 \tau$. This condition upper bounds the CSL parameter $\gamma^0_{\mathrm {CSL}}$ as
\begin {equation}
\gamma^0_{\mathrm {CSL}} < \bar {\gamma}_{\mathrm {CSL}} = \frac{4 m_0^2 a_{\mathrm {CSL}}}{10 \tau m^2 \sigma^2_{\mathrm {max}}  f(R/a_{\mathrm{CSL}})}.
\end {equation}

\begin{figure}[!t]
\begin {center}
   \includegraphics[width=\columnwidth]{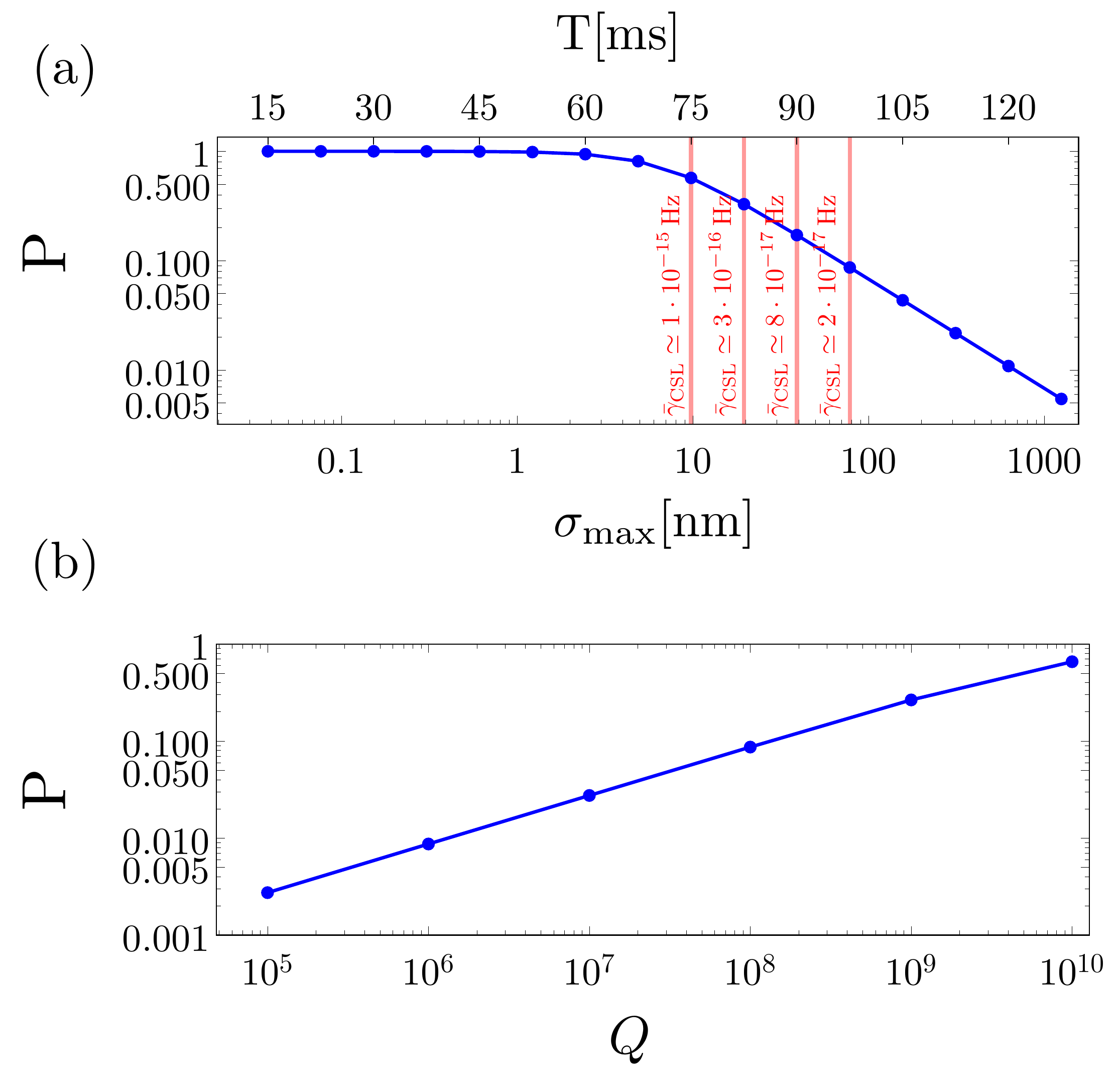} 
\end{center}
\caption {A single particle of radius $R_0=250$~nm is prepared in the ground state and periodically undergoes a sequence of frequency shifts between the ideal values $\omega_1=(2\pi)\  100$~Hz and $\omega_2=2 \omega_1$. (a) Purity at the end of a protocol of duration T and corresponding maximum spreading $\sigma_{\mathrm {\max}}$, in the presence of a thermal bath with $\bar n= 100$ phonons and coupled to the system at a rate $\Gamma=\omega_1/Q$, with quality factor of $Q=10^8$.  On the  vertical lines we display the parameter $\bar \gamma_{\mathrm {CSL}}$ which would predict, for the corresponding spreading, a coherence time one order  of magnitude larger than $4/3 \tau_1$. (b) Purity as a function of the quality factor for a protocol of duration $\tau \simeq 100$~ms with maximum spreading $\sigma_{\mathrm {max}}\simeq 80$~nm.} 
\label {fig2:sup}
\end{figure}

The experimental certification of coherence at the end of our protocol for growing values of $\sigma_{\mathrm {max}}$ would set tighter and tighter bounds on $\gamma^0_{\rm CSL}$. We display in Fig.~(\ref{fig2:sup}) the purity retained after the squeezing and unsqueezing sequences in the presence of dissipation due to a thermal environment. In panel (a), we show the value of the purity as a function of the total duration of the protocol and the maximum spreading $\sigma_{\mathrm {max}}$ reached at the middle of the protocol. As expected, we observe how the purity decreases with increasing protocol length and squeezing. We indicate the approximated bound $\bar \gamma_{\mathrm {CSL}}$ that reaching such a delocalization would set. In  panel (b), in order to explore the role of environmental decoherence, we display the purity as a function of the quality factor for a protocol of maximum spreading $\sigma_{\mathrm {max}}\simeq 80\, \mathrm {nm}$ for a particle with a mass of $m\simeq 10^{-16 }$ kg (corresponding roughly to $10^{10}$ nucleons), which would set a bound of $\bar \gamma_{\mathrm {CSL}}\simeq 2 \cdot  10^{-17 }$ Hz.

 It is worth to mention, that the impact of background gas molecules with the  levitated particle will destroy the superposition, thus hindering the test of collapse models. Specifically, the collision rate with air molecules on a spherical particle of radius $R$ can be approximated as ~\cite{ChangRP+2010}
\begin{equation}
\label{collrate}
R_{\rm air} \approx \frac{\pi \bar v P R^2}{k_{B} T},
\end{equation}
where $P$ is the chamber pressure, $\bar v = \sqrt{\frac{3 K_B T}{m_a}}$ the average velocity of the air molecules with $m_a \sim 28.97$ amu their mass. For the particle sizes considered here, $R= 250\ $nm, and at temperatures of $T \sim 100$ K, we find a collision rate $R_{\rm air}\approx  10^{10} P$ [Hz]. This implies that we require ultra-high vacuum below $10^{-10}$ Pa in order to get a collision rate $\sim 1$ Hz.


\end{document}